# Mass spectrometric analyses of high performance polymers to assess radiopurity as ultra low background materials for rare event physics detectors

Jay W. Grate,* Isaac J. Arnquist,* Eric W. Hoppe, Mary Bliss, Khadouja Harouaka, Maria Laura di Vacri, and Sonia Alcantar Anguiano

Pacific Northwest National Laboratory, P.O. Box 999, Richland, WA 99352

**Abstract**

The mass concentrations of $^{232}$Th and $^{238}$U in several strong, unfilled, high performance polymers are reported as measures of their radiopurity. Highly radiopure polymers are required as dielectric materials in the construction of rare event physics detectors, in order to minimize background signals arising from the detector materials themselves. New data are reported for carefully sourced samples of polyetheretherketone (PEEK), polyetherketoneketone (PEKK), a polyamide imide (PAI, branded as Torlon) and polybenzimidazole (PBI). Data for solid polyetherimide (PEI) are also discussed and new data for PEI in the form of a commercial 3D printing filament stock are reported. Strong high performance polymers PEKK, PBI, PAI and PEI were found with levels for $^{232}$Th and $^{238}$U that are below one mBq/kg, including the PEI 3D printing filament. Specifically, for $^{232}$Th and $^{238}$U respectively, in μBq/kg (emphasis "micro"Bq/kg), we found values of 149 and 184 for Arkema Kepstan PEKK 6002 flake; 69 and 2250 for Solvay Ketaspire PEEK flake; 346 and 291 for PBI Performance Products low metals grade 100 mesh PBI powder; 66 and 105 for Drake Plastics PAI Torlon T-4200 pellets; 401 and 285 for Drake Plastics cured PAI rod; 32 and 41 for Ensinger PEI Ultem 1000 solid; and 15 and 85 μBq/kg, for ThermaX PEI Ultem 1010 filament material. These results were all obtained using a novel dry ashing method in crucibles constructed of ultra low background (ULB) electroformed copper. Samples were spiked with $^{229}$Th and $^{233}$U as internal standards prior to ashing and determinations were made by inductively coupled plasma mass spectrometry (ICP-MS). Radiopurity is displayed graphically relative to numerical measures of mechanical strength for these and several other polymers.

* Jay W. Grate: jwgrate@pnnl.gov, 509-371-6500

*Isaac J. Arnquist: isaac.arnquist@pnnl.gov, 509-372-6531

# 1 Introduction

The detection of rare events is at the heart of many fundamental physics questions that probe the nature of the world in which we live. Rare event physics seeks to explore and understand matter-antimatter asymmetry, neutrino physics, particle physics, dark matter, and the composition of the universe. In the search for dark matter, solar neutrinos, and neutrinoless double beta decay, particle interactions with a scintillator or a crystal are transduced into signals. The backgrounds of many such detectors must be rigorously minimized. Therefore, the materials of construction will ideally have little to no radioactivity of their own, which would otherwise generate signals obscuring the rare event signal of interest. Materials and their assay have been reported in papers related to a host of detectors in major physics projects[1-18]. Each generation of dark matter detector requires more sensitivity than previous efforts. Thus, the radiopurity requirements for component materials become ever more stringent, and hence the need for efficient assay methods capable of detection to lower and lower concentrations.

In the present paper, we focus on the determination of $^{232}$Th and $^{238}$U at very low levels in strong high-performance polymer materials, using inductively coupled plasma mass spectrometry (ICP-MS). To obtain low process blanks and low detection limits, we use electroformed copper foil as an ultra low background (ULB) material to make thin walled containers serving as crucibles for dry ashing. This metal, with purity levels in bulk electroformed samples below 8.4 and 10.6 fg/g (ppq) for $^{232}$Th and $^{238}$U, respectively[11], is a preferred construction material in ULB rare event detectors. These unusual low-mass, high-purity, crucibles are completely dissolved along with the ash prior to analysis. This approach is used for all polymers in this report, and additional results using other methods such as microwave digestion are reported for comparison in specific cases.

ICP-MS is an efficient and increasingly necessary instrumental technique to evaluate materials that are of high radiopurity. While gamma spectroscopy has been used extensively in the past, it lacks the ultimate sensitivity to quantify low levels, even when using large sample masses counted for weeks or months on ULB detectors. When gamma spectroscopy is sufficiently sensitive, it has the benefit that the activity of different isotopes in the decay sequences of $^{232}$Th and $^{238}$U can be assessed for secular equilibrium. Nuclear activation analysis (NAA) can achieve detection of very low levels[10, 12, 14, 19]; however the cumbersome process requires expensive reactor time, and is much more time-consuming than the dry or wet ashing of polymer samples and dissolution for ICP-MS.

It is a convention in ULB physics to reference radiopurity to the radioactivity concentrations of the parent isotopes that are the heads of major decay chains, *i.e.*, to $^{232}$Th and $^{238}$U. Even when measuring radiopurity using gamma spectroscopy and measuring the gamma rays of daughter isotopes (*e.g.,* $^{208}$Tl, $^{226}$Ra) in the decay chain, results are conventionally reported as activities of $^{232}$Th and $^{238}$U and assume secular equilibrium from parent to progeny. Radioactivity concentrations of $^{232}$Th and $^{238}$U are input into complex detector models to understand detector backgrounds and simulate sensitivity reach. Models take into account the entirety of decay chain isotopes. Mass concentrations of $^{232}$Th and $^{238}$U determined by mass spectrometry are converted to the activity concentrations by isotope specific conversion factors, *i.e.*, 1 pg/g of $^{232}$Th and $^{238}$U are equivalent to 4.1 and 12.4 μBq/kg, respectively. These conversion factors can be calculated from the specific activity and the atomic mass. Assay data in this paper are reported in both the mass concentrations, pg/g, and the radioactivity concentration as μBq/kg.

Publicly-available radiopurity-relevant isotopic data on contaminants in some unfilled and filled plastic materials are provided in **Table 1** in mBq/kg units (emphasis 'milli'Bq/kg, compared to the 'micro'Bq/kg levels we will be presenting in the remainder of the paper)[20]. Values were determined by gamma spectroscopy as part of the development of the SuperCDMS detector for weakly interacting massive particles (WIMPs - a candidate particle to describe dark matter)[21, 22]. Two samples of PE are listed. Three examples of liquid crystal polymer (LCP) containing fillers are listed; these illustrate the poor radiopurity of filled plastics, with the distribution of contaminants dependent on the nature of the filler. The presence of fillers leads to contaminant levels 3 or 4 orders of magnitude higher than ideal plastic materials that have levels below a single mBq/kg. Unfilled samples of Zytel-branded nylon polyamide, a Vespel-branded polyimide, polyethersulfone (PES), and polyetheretherketone (PEEK) are also listed: these are typically stronger plastics than PE, but with regard to radiopurity they have values equal to or greater than one mBq/kg for the $^{238}$U and/or $^{232}$Th decay chains. A compilation of assay data on plastics from multiple journal articles was previously provided in the Supplementary Information in previous work[23].

**Table 1**. Gamma spectroscopic assay on several plastics associated with the development of the SuperCDMS detector.[a]

| | $^{238}U$ | $^{232}Th$ |
|---|---|---|
| | mBq/kg | mBq/kg |
| Polyethylene (PE) [b] | 1.836 | 0.563 |
| Polyethylene (PE) [c] | 1.85 | 5.99 |
| Zytel nylon polyamide | < 1.34 | 1.644 |
| Polyethersulfone, PES | 1.208 | 1.141 |
| Vespel polyimide | 7.72 | 9.95 |
| Polyetheretherketone (PEEK) [d] | 20.53 | 11.85 |
| Liquid crystalline polymer (LCP)/ Glass-filled | 8234 | 10990 |
| Liquid crystalline polymer (LCP)/ Mica-filled | 380.7 | 73.79 |
| Liquid crystalline polymer (LCP) Talc-filled | 2788 | 29.38 |

[a] - Radiopurity.org retrieved September 7, 2019.
[b] - CDMS poly arcs
[c] - 8 CDMS inner shield layer poly arcs
[d] - Specifically, Victrex 450G903B, as black pellets.

Here we focus on the radiopurities of unfilled, unpigmented, high performance polymers, seeking both high strength and high purity. The chemical structures of five such polymers to be evaluated in this study are shown in **Figure 1**. These are highly aromatic polymer structures. These include two types of polyaromaticetherketones (PAEKs), PEEK and polyetherketoneketone (PEKK). Additional high strength materials included are PEI, polyamide imide (PAI, branded as Torlon), and polybenzimidazole (PBI). By high performance, we are emphasizing primarily mechanical strength, although other properties such as low coefficient of thermal expansion are also favorable since many detectors are cooled to cryogenic temperatures. Our assay data demonstrate that a number of these high performance polymer materials can be sourced that have radiopurities at or below one mBq/kg. To show polymers with both high purity and high strength, as assessed by an objective criterion, we have plotted these and prior assay results against mechanical properties. We found that PEI is a particularly clean high performance polymer as solid; additionally, we found that PEI available commercially in filament for 3D printing by fused deposition modeling (FDM), a form of additive manufacturing, is also of noteworthy radiopurity.

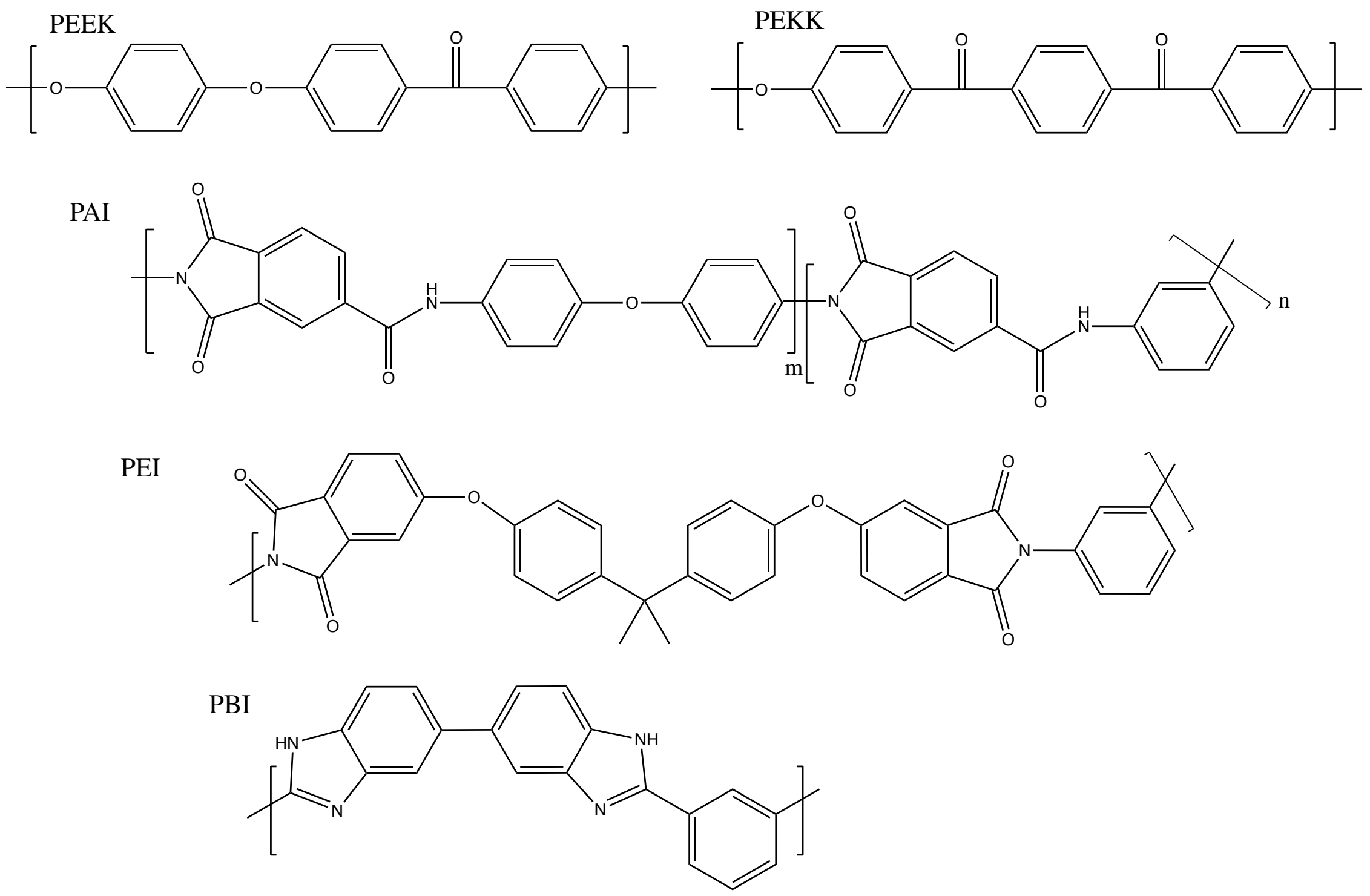

**Figure 1**. Chemical repeat unit structures of some high performance polymers assayed and/or discussed in this paper. Exact PAI structure depends on the grade of Torlon; typically m=0.7, n=0.3 for the 4200 series[24, 25].

# 2 Experimental

## 2.1 Materials

PEEK in the form of Ketaspire KT-880P flake was kindly provided by Solvay Specialty Polymers, Kernersville, NC. Two PEKK resins were ordered from Arkema as flake material: Kepstan 6002PF 60/40 PEKK - Med Flow Reactor Flake, and Kepstan 7002PF 70/30 PEKK - Med Flow Reactor Flake. Flake materials are powders recovered from synthesis reactors, but they are not free-flowing, they are aggregated into "flakes". PBI U60SD 100 mesh powder was ordered from PBI Performance Products, Inc Charlotte, NC, and arrived labeled on the bottle as "low metal". PAI Torlon T-4200 was kindly provided as samples by Drake in pellet and cured rod; the cured rod was prepared from the same batch number of resin as the pellets. PAI rod was subsampled by cutting with clean metal tools. ThermaX Brand PEI 3D Filament in 1.75 mm diameter was purchased from 3DXTech Advanced Materials; this material is made from Ultem 1010 resin. PEI solid material was assayed previously at PNNL and published[26]. This prior solid sample was derived from Ultem 1000 resin[26].

## 2.2 Sample preparation

A Class 10,000 cleanroom at Pacific Northwest National Laboratory (PNNL) and a laminar flow hood providing a Class 10 environment were used for sample preparations and anion exchange separations. As most of the screened materials were powder or flake raw materials, or pellet, there was no need to subsample and, then, further clean the material before analysis. For PAI solid (rod), however, the material was cut into smaller subsamples using a new, cleaned wire cutter. In order to minimize contamination from subsampling, the subsampled PAI rod was sonicated in pure ultraclean ethanol for 20 mins followed by a minor leach in 5% $HNO_3$ for 20 minutes to remove surficial contaminants from subsampling. The subsamples were triply rinsed in DI water and allowed to air dry in a class 10 laminar flow hood before assay. For PEI (Ultem) filament, the material was cut into smaller subsamples using cleaned stainless steel scissors. Subsamples were sonicated in 2% micro-90 detergent solution for 10 minutes, triply rinsed in DI water, sonicated in Optima grade 6M $HNO_3$ for 20 minutes, triply rinsed in DI water and allowed to air dry in a class 10 laminar flow hood before assay.

## 2.3 Polymer Decomposition Methods

This study utilized three polymer decomposition methods in order to remove the $^{232}$Th and $^{238}$U contaminants from the polymer matrix previous to solution nebulization analysis via ICP-MS. The three methods used include dry ashing by ULB EF-Cu crucibles for all polymers, dry ashing in quartz crucibles for PBI and PEI, and, as well as wet ashing using microwave digestion for PAI and PEI.

### 2.3.1 Dry Ashing Crucible Preparations and Recovery

For dry ashing with quartz crucibles, the subsamples were placed in acid leached and validated cleaned quartz crucibles along with a known amount of $^{229}$Th and $^{233}$U tracer solution. Process blanks (empty quartz crucibles) were spiked with tracer and carried through the process as well. Samples were submitted to the dry ashing method described in the following section. Post-ashing residual analytes/tracer were recovered from the quartz crucibles by boiling off 2 mL of 8M $HNO_3$ aliquots before reconstitution in 1.8 mL 2% $HNO_3$ for analysis via ICP-MS.

Dry ashing in ULB EF-Cu crucibles followed the methods described in detail previously [23]. Briefly, small mass crucibles of ULB EF-Cu foils were folded into a "boat" shape to hold up to 100s of mg of polymer. The crucibles were lightly etched with nitric acid to remove any superficial contamination from handling after crucible formation. ULB EF-Cu crucibles were supported on cleaned, validated quartz holders and then loaded with polymer subsamples and spiked with a known amount of $^{229}$Th and $^{233}$U tracer solution. Process blanks (empty ULB EF-Cu crucibles) were spiked with tracer and carried through the process as well. Post-ashing residual analytes/tracer were recovered by digesting the entirety of the ashed ULB EF-Cu in 8M $HNO_3$ and then separating the U and Th from the Cu matrix following the adapted method described here[11, 23]. Elution volumes were analyzed via ICP-MS.

### 2.3.2 Dry Ashing Program

Dry ashing using both ULB EF-Cu and quartz crucibles utilized a programmable TransTemp quartz tube furnace (Thermcraft Inc., Winston-Salem, NC). Samples were ashed using a slow-ramping heating program with a maximum temperature of 800 $^{o}$C in the presence of air at a flow rate of 4 L/min. Typical heating runs took 10-14 hours overnight before reclaiming the ashed crucibles the following day. Polymers were fully decomposed after the heating program.

### 2.3.3 Wet Ashing Using Microwave Digestion

A Mars 6™ microwave digestion system with iPrep™ vessels (CEM Corp., Matthews, NC) was used for microwave-assisted digestion of PAI and PEI. Samples were loaded into cleaned and validated vessels, along with a known amount of $^{229}$Th and $^{233}$U tracer solution, and 5 mL of Optima Grade $HNO_3$ (Fisher Scientific). Process blanks were carried through the procedure; vessels were filled with tracer solution and the concentrated nitric acid. The heating program used a 30 minute heat ramp to 250$^{o}$C and a hold at this temperature for 30 minutes before cooling to room temperature. Samples were retrieved, transferred to cleaned/validated perfluoroalkoxy alkane (PFA) vials (Savillex, Eden Prairie, MN), and reconstituted into 1.8 mL of 2% $HNO_3$ before analysis via ICP-MS.

## 2.4 ICP-MS Instrumentation

Samples were analyzed for Th and U using an Agilent 8800 or 8900 series ICP-MS (Agilent Technologies, Santa Clara, CA) with an integrated autosampler and PFA microflow nebulizer (Elemental Scientific, Omaha, NE).

# 3 Results and discussion

## 3.1 Sample preparation approach

In analyzing for trace contaminants in plastic materials, the contaminant atoms must be transferred from the matrix of the plastic material into a solution matrix for isotope dilution (ID)-ICP-MS with liquid nebulization sample introduction. This must be carried out while minimizing new contaminants being introduced to the sample from the laboratory environment, the processing method, or the containers. We recently introduced an ultra trace method for ashing polymers; it entails dry ashing the sample in containers that are formed by folding foil pieces of ULB EF-Cu into boats that serve as crucibles (**Figure 2**)[23, 26]. After ashing, the entire foil crucible and the ash are dissolved in acid and sent through a Cu removal anion exchange column separation process[11]. Process blanks are consistent and low, indicating that the ULB EF-Cu is a consistent material, and the process is repeatable in our clean work environment. We continue to obtain absolute detection limits, calculated as 3 x standard deviation of the process blanks, in the range of typically 20-100 fg within a sample set[23].

We used tracers of $^{229}$Th and $^{233}$U as internal standards for quantification[11, 27], spiking all process blanks and samples[23, 26]. In this method, tracer recoveries are high and correspond to the recoveries of the separation, *i.e.*, typically *ca.* 80% for $^{229}$Th and 90% for $^{233}$U. All polymers assayed in this paper (PEEK, PEKK, PBI, PAI, and PEI) were determined using the copper crucible ashing method and the tracers were added prior to dry ashing. The tracer recovery numbers were consistent with what we've reported previously for other samples and process blanks, *i.e.*, *ca.* 80% for $^{229}$Th and 90% for $^{233}$U. In a few cases, assays were also conducted with dry ashing in quartz crucibles, or wet ashing by microwave digestion, and compared. In these cases, the containers are not dissolved with the ash, and no subsequent column separation is needed or performed prior to mass spectrometric determination.

These procedures with low mass ULB containers enable very sensitive determinations; the use of internal standards and observation of high tracer recoveries lends confidence to the accuracy of the assay results[23]; and the approach may be regarded as a primary ratio method of measurement[28-34].

Polymer samples of PEEK flake, PAI Torlon pellets and rod, and PEI solid and filament, are shown in **Figure 2, 3,** and **4,** respectively.  The "imide" polymers are brown or amber due to charge transfer interaction among donor and acceptor structures in the material.

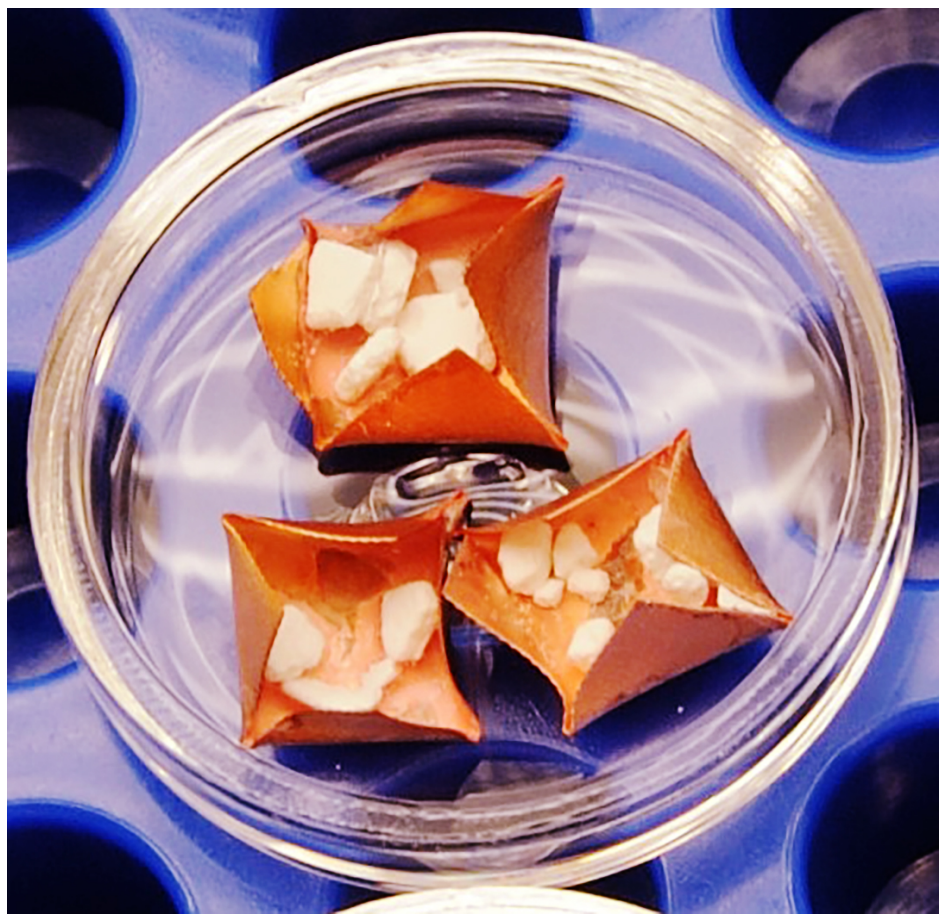

**Figure 2**.  Samples of PEEK flake in ULB EF-Cu crucibles prior to dry ashing.

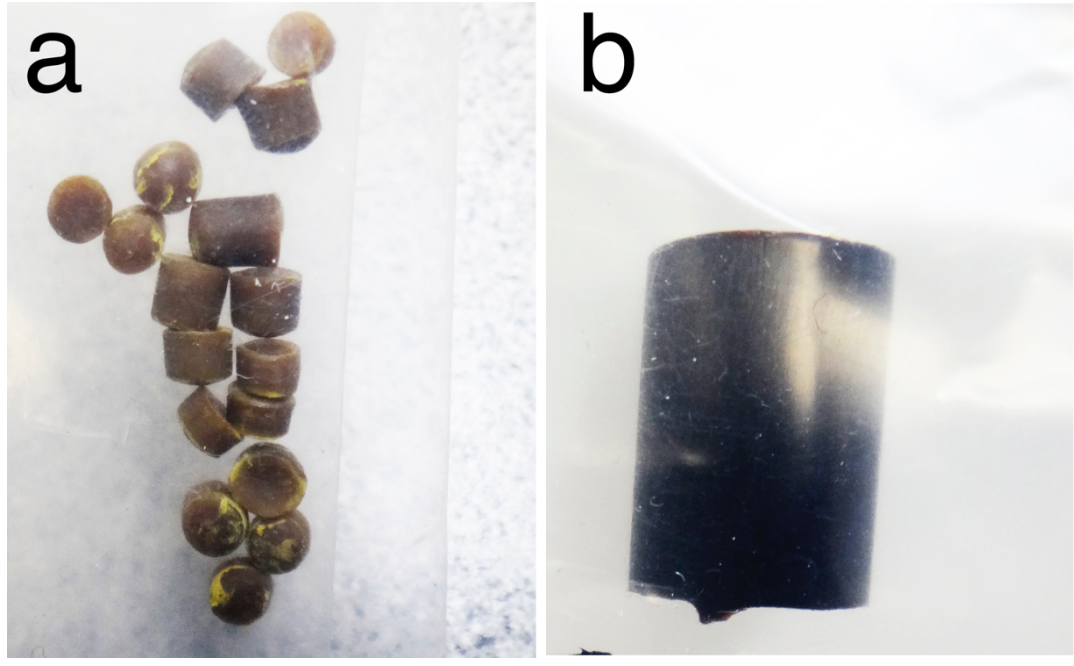

**Figure 3.** Images of PAI Torlon T-4200 samples:  a) pellets as received double-bagged, and b) cured rod piece in bags.  The pellets and rod are 2mm and 10 mm in diameter, respectively.

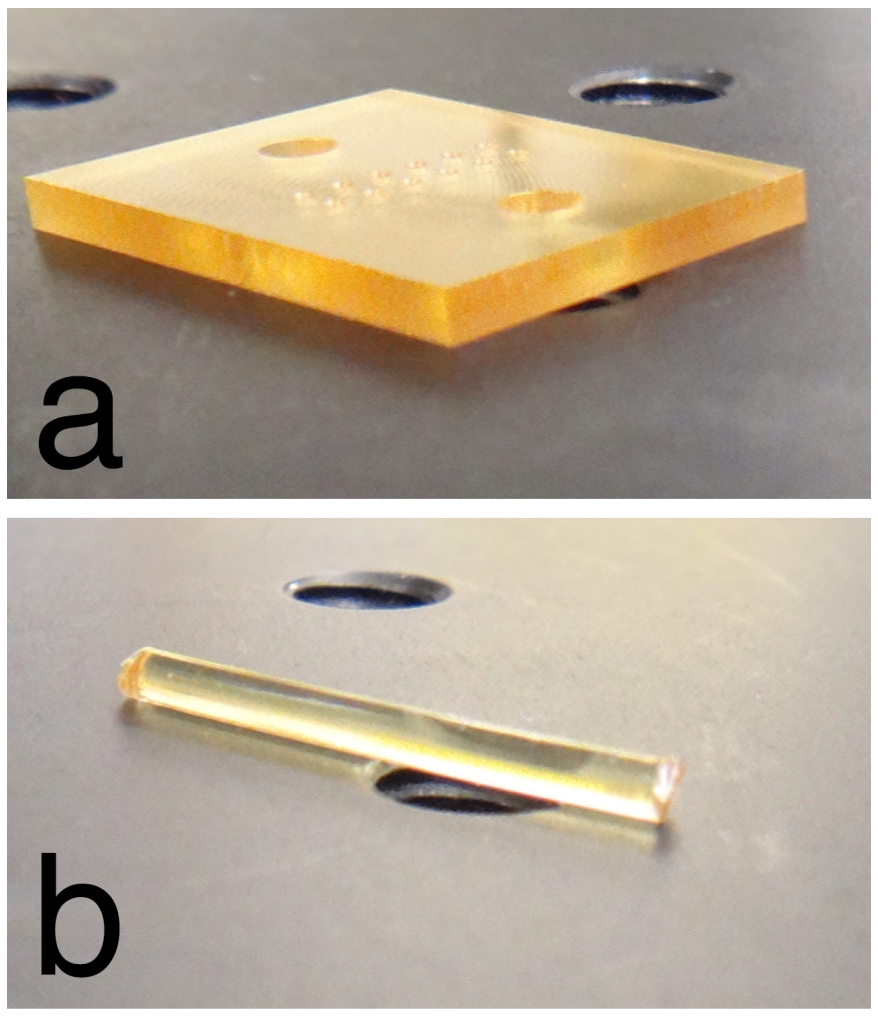

**Figure 4**. Images of PEI material forms: a) solid Ultem 1000 in the form of a machined prototype connector part, and b) Ultem 1010 in filament form for 3D printing. The filament is 1.75 mm in diameter.

## 3.2 Poly aromatic ether ketones

PAEKs are strong, dimensionally stable plastics. We assayed flake samples of PEEK and PEKK samples, using the dry ashing method with ULB EF-Cu, and results are provided in **Table 2**. We previously assayed PEKK from a different source and obtained values in the thousands of μBq/kg for $^{232}$Th and $^{238}$U[27], indicating overall the poor radiopurity of the material in that prior study.

**Table 2. Determination of $^{232}$Th and $^{238}$U in PEEK and PEKK Flake Samples**

| | | | $^{232}$Th | | $^{238}$U | |
|---|---|---|---|---|---|---|
| **Polymer** | **#** [a] | **Method**[b, c] | **pg/g** | **μBq/kg** | **pg/g** | **μBq/kg** |
| PEEK_flake | 1 | EF-Cu | 13.4 | 55.0 | 187 | 2310 |
| | 2 | | 23.4 | 96.0 | 181 | 2240 |
| | 3 | | 13.4 | 54.9 | 177 | 2190 |
| PEKK 7002_flake | 1 | EF-Cu | 309 | 1266 | 261 | 3240 |
| | 2 | | 453 | 1857 | 664 | 8240 |
| | 3 | | 384 | 1576 | 410 | 5080 |
| PEKK 6002_flake | 1 | EF-Cu | 32.3 | 133 | 13.0 | 161 |
| | 2 | | 30.1 | 123 | 14.5 | 180 |
| | 3 | | 45.7 | 187 | 16.9 | 210 |

[a] Number within sample set.
[b] Tracer addition Pre- dry ashing. EF-Cu = dry ashing in copper boat crucibles
[c] Sample masses were in the range of 20-45 mg

The PEEK flake material we sourced for the current study yielded assay results averaging 69 and 2250 µBq/kg for $^{232}$Th and $^{238}$U. While the value for $^{238}$U exceeds our goal for values below one mBq/kg, it is still substantially more pure than Victrex 450G903B PEEK pellets assayed for the SuperCDMS project; values on radiopurity.org are reported as 11,850 and 20,530 µBq/kg for $^{232}$Th and $^{238}$U[20]. Our values are also much lower than data reported for a PEEK sample analyzed for the NEXT-100 double beta decay experiment, *i.e.*, 14,900 µBq/kg for $^{232}$Th. [5]. A sample of Victrex PEEK assayed for the Majorana project by NAA indicated values of less than 1,600 and 63,000 µBq/kg for $^{232}$Th and $^{238}$U respectively[2]. PEEK fasteners assayed for the XENON1T by counting indicated values of less than 20,000 and 56,400 µBq/kg for $^{228}$Ra and $^{238}$U respectively[16]. Ketaspire PEEK is produced by a different process than that used to produce commodity Victrex PEEK, and is the basis material for a biomedical grade of PEEK.

The 7002 PEKK material yielded assay results averaging 157 and 5520 µBq/kg for $^{232}$Th and $^{238}$U. However, the 6002 resin PEKK material yielded assay results averaging 149 and 184 µBq/kg for $^{232}$Th and $^{238}$U. The PEKK 6002 resin is substantially purer than either the PEKK in our prior study[27], or the PEKK 7002 in this study. The PEKK is also purer than the Ketaspire PEEK flake or other PEEK materials from the literature. The PEKK 6002 resin falls well within our goal, and provides very good radiopurity for a strong polymer material.

The results for the PEKK resins illustrate the benefits of having a rapid analysis method with a low detection limit that enables screening, and the fact that even closely related samples, *i.e.*, 6002 and 7002 PEKK resins, can vary significantly in purity. The $^{238}$U levels in PEKK 6002 are over an order of magnitude lower than in PEKK 7002. If only a single example of PEKK had been assayed, the cleaner material might not have been found.

## 3.3 Polybenzimidazole powder

PBI is among the strongest of all unfilled plastic polymers, with outstanding compressive strength and low coefficient of thermal expansion[35, 36]. A grade of PBI has been developed for semiconductor industry applications where the powder undergoes an extractive metals removal process. We obtained low metals grade PBI as 100 mesh powder for assay. Results are provided in **Table 3**.

**Table 3. Determination of $^{232}$Th and $^{238}$U in PBI Powder Samples**

| | | | $^{232}$Th | | $^{238}$U | |
|---|---|---|---|---|---|---|
| **Polymer** | **#** [a] | **Method**[b, c] | **pg/g** | **µBq/kg** | **pg/g** | **µBq/kg** |
| PBI powder | 1 | EF-Cu | 89.7 | 368 | 19.9 | 247 |
| | 2 | | 97.0 | 398 | 18.8 | 234 |
| | 3 | | 66.4 | 272 | 31.6 | 392 |
| PBI powder | 1 | Quartz | 172 | 706 | 16.0 | 198 |
| | 2 | | 144 | 592 | 19.0 | 236 |
| | 3 | | 151 | 620 | 19.2 | 238 |

[a] Number within sample set.
[b] Tracer addition Pre- ashing. EF-Cu = dry ashing in copper boat crucibles, Quartz = dry ashing in quartz crucibles.
c Sample masses were in the range of 40-100 mg

Using the copper crucible dry ashing method, values for $^{232}$Th and $^{238}$U averaged 346 and 291 µBq/kg, respectively. By contrast, using the quartz crucible method for dry ashing, our assays yielded average values of 639 and 224 µBq/kg for $^{232}$Th and $^{238}$U, respectively. The $^{232}$Th value is notably higher than that obtained from the Cu-boat method, while the $^{238}$U value is in reasonable agreement. In the quartz crucible method, the tracer recoveries averaged 63% and 86% for $^{229}$Th and $^{233}$U, respectively. We have previously noted challenges with dry ashing polymer samples using quartz crucibles[27]. In the present case, tracer recovery for $^{229}$Th was observed to be low. If the added Th tracer were adsorbed to the quartz material, and not recovered in the same proportions as $^{232}$Th from within the polymer, this would lead to values biased higher for the $^{232}$Th assay. $^{232}$Th values in **Table 3** for PBI are higher in the quartz crucible samples than using the Cu crucible method.

## 3.4 Imide-containing aromatic polymer solids

PAI and PEI both have imide-containing molecular structures (**Figure 1**); samples are shown in **Figure 3** and **Figure 4**. The amber to brown color of such materials arises from charge transfer interactions.

**Table 4. Determination of $^{238}U$ and $^{232}Th$ in PAI Pellet and Rod Samples**

| | | | $^{232}Th$ | | $^{238}U$ | |
|---|---|---|---|---|---|---|
| **Polymer** | **#** [a] | **Method**[b, c] | **pg/g** | **μBq/kg** | **pg/g** | **μBq/kg** |
| PAI pellet | 1 | EF-Cu | 13.1 | 53.7 | 8.56 | 106 |
| | 2 | | 18.9 | 77.6 | 8.33 | 103 |
| PAI rod [d] | 1 | MW | 140 | 575 | 37.5 | 465 |
| | 2 | | 84.2 | 345 | 25.9 | 321 |
| | 3 | | 125 | 512 | 25.8 | 320 |
| PAI rod [e] | 1 | MW | 58 | 238 | 38 | 471 |
| | 2 | | 117 | 481 | 34.8 | 432 |
| | 3 | | 74.1 | 304 | 24.5 | 303 |
| PAI rod [e] | 1 | EF-Cu | 85.9 | 352 | 23.5 | 292 |
| | 2 | | 140 | 574 | 27.5 | 340 |
| | 3 | | 67.3 | 276 | 18.0 | 223 |

[a] Number within sample set.
[b] Tracer addition Pre- ashing. MW = microwave digestion as a wet ashing method. EF-Cu = dry ashing in copper boat crucibles
[c] Sample masses were in the range of 15-120 mg
[d] Normal cleaning
[e] Rigorous cleaning, see text.

PAI is an interesting high performance polymer that can be cured after solid shapes have been formed by melt-processing methods[37]. As a melt-processable material, it is described as having the highest strength and lowest coefficient of thermal expansion of thermoplastics. We obtained Torlon T-4200[37, 38] in pellet and thermally cured rod forms.(**Figure 3**) It should be noted that another grade, T-4203, is often labeled as an unfilled plastic, but it actually has a few percent of white titanium dioxide added to obtain a more pleasing gold color from this otherwise amber brown resin. The T-4200 we sourced does not have added titanium dioxide. Unlike the PBI, PEEK, and PEKK, we obtained assay results from solids rather than powder raw materials; results are in **Table 4**.

Unlike some polymers, PAI can be wet ashed with acids, and we initially attempted to assay PAI pellets after microwave digestion. No positive results above backgrounds were obtained for the pellets. With remaining pellet sample, we used the copper crucible dry ashing method, yielding assay values averaging 66 and 105 μBq/kg for $^{232}Th$ and $^{238}U$, respectively. Samples of PAI rod material did provide positive values by microwave digestion. These assays were carried out after cleaning the rod subsamples using *ca.* 20 minute sonications in ultrapure ethanol followed by dilute nitric acid (5%). A more rigorous cleaning was also tested, which involved the aforementioned steps as well as very brief leaches (*ca.* 1 min) in concentrated $HNO_3$ and HCl. The extra cleaning made little difference. Values of 477 and 341 μBq/kg were obtained for $^{232}$Th, while values of 369 and 402 μBq/kg were obtained for $^{238}$U. (For the microwave method for PAI, the tracer recoveries were at or near 100%.) The additional assay of rod subsamples by the Cu crucible method, also using the more rigorous cleaning procedure, yielded values of 401 and 285 μBq/kg for $^{232}$Th and $^{238}$U, respectively. Regardless of sample preparation, the cured rod material has the same overall order of magnitude of radiopurity in the low hundreds of μBq/kg for each element. The extruded and cured rod material is not as clean as the raw pellet material. The unpigmented Torlon 4200 PAI pellet is a high performance polymer with very good radiopurity, well below our upper target of one mBq/kg. PAI of Torlon 4203 and 4203L have been previously assayed for the XENON1T radioassay program; in measurements made by ICP-MS, values of 490 and 4900 μBq/kg were reported for $^{232}$Th and $^{238}$U, respectively. [16].

PEI is an amorphous amber-colored thermoplastic with excellent dimensional stability and high mechanical strength. Solid and filament forms of PEI are shown in **Figure 4**. It has been said to be "*strong enough to replace steel in some applications and light enough to replace aluminum in others*"[39]. We previously assayed solid PEI material derived from Ultem 1000, obtaining values averaging 32 and 41 μBq/kg for $^{232}$Th and $^{238}$U, respectively, using the Cu crucible method[26]. This is a particularly clean material among the high performance polymers discussed here. Ultem 1000 resin is the standard unfilled grade of PEI.

Here we present new data for commercial PEI filament material that is marketed for 3D printing by fused deposition modeling (FDM), a form of additive manufacturing. This material was assayed using all three methods of sample decomposition (Cu crucible, quartz crucible and microwave). Results are reported in **Table 5**. Using the copper crucible dry ashing method, values for $^{232}$Th and $^{238}$U averaged 15 and 85 μBq/kg, respectively. Similarly, using the microwave digestion method for wet ashing, our assays yielded average values of 21 and 87 μBq/kg for $^{232}$Th and $^{238}$U, respectively. Copper crucible and microwave methods provide very similar results providing evidence of method validity. The results are similar to those of the solid PEI (Ultem 1000) assayed previously in the respect that the Ultem 1010 filament is also very radiopure.

**Table 5. Determination of $^{238}$U and $^{232}$Th in PEI Filament Samples**

| | | | $^{232}$Th | | $^{238}$U | |
|---|---|---|---|---|---|---|
| **Polymer** | **#** [a] | **Method**[b, c] | **pg/g** | **µBq/kg** | **pg/g** | **µBq/kg** |
| PEI filament | 1 | EF-Cu | 4.53 | 18.6 | 8.65 | 107 |
| | 2 | EF-Cu | 3.12 | 12.8 | 6.06 | 75.1 |
| | 3 | EF-Cu | <3.04 | <12.5 | 5.93 | 73.5 |
| PEI filament | 1 | MW | 2.98 | 12.2 | 6.73 | 83.4 |
| | 2 | MW | 6.41 | 26.3 | 7.39 | 91.6 |
| | 3 | MW | 5.96 | 24.5 | 7.02 | 87.1 |
| PEI filament | 1 | Quartz | 24.1 | 99.0 | 45.3 | 561 |
| | 2 | Quartz | 9.26 | 37.9 | 7.68 | 95.3 |
| | 3 | Quartz | 13.2 | 54.2 | 10.4 | 129 |

[a] Number within sample set.
[b] Tracer addition Pre- ashing. EF-Cu = dry ashing in copper boat crucibles. Quartz = dry ashing in quartz crucibles. MW = microwave digestion as a wet ashing method.
[c] Sample masses were in the range of 45-215 mg

Using the quartz crucible method for dry ashing, our assays yielded somewhat higher average values of 64 and 262 µBq/kg for $^{232}$Th and $^{238}$U, respectively. The tracer recoveries were near 100% for both $^{229}$Th and $^{233}$U. Nevertheless, we have previously noted challenges with dry ashing polymer samples using quartz crucibles[27]; this is a strong motivation for using ULB EF-Cu copper crucibles.

## 3.5 Mechanical properties versus radiopurity

The prior sections have provided numerical data on the contaminants in the polymer samples, while qualitative statements have been made about materials strength. In this section we visually illustrate how polymers compare in terms of both mechanical strength and radiopurity, using ASTM standard test data for numerical indicators of strength. We compiled this information for polymers we have assayed in this paper and in prior papers. We use the $^{238}$U levels as an indicator of radiopurity; when multiple data sets for a given polymer type were available, the most favorable radiopurity values are selected as an indication of what is possible. For example, for PEKK we plotted the data for the 6002 resin, and for PEI we used the data for the solid form.

**Figs. 5** presents tensile strength as an example. Radiopurity is represented as the concentration of $^{238}U$ on the vertical axis in log units, with radioactivity concentration on the left and mass concentration on the right (pg/g corresponds to parts per trillion). The optimal location in these plots is toward the lower right corner, where purity and strength are both high. All the high performance polymers in this paper are beyond the half way point from left to right on the plots indicating strength; and four of the five have radiopurities with $^{238}U$ below one mBq/kg. PEI appears to be best among the high performance polymers assayed thus far, where strength and dimensional stability are needed in addition to purity. Polymers of excellent radiopurity used in the past, such as PTFE and PCTFE, [26, 40, 41 , 42, 43] reside in the lowest quarter of the plots with regard to strength.

Where a given polymer was available in multiple forms, *e.g.*, raw material as powder, or pellets, or solid, a vertical line has been drawn to connect them (*i.e.*, PVDF and PAI, and the two forms of PEI). It is apparent that solids are less pure than pellets or powder, apparently picking up contamination when transformed from raw materials to other solid forms.

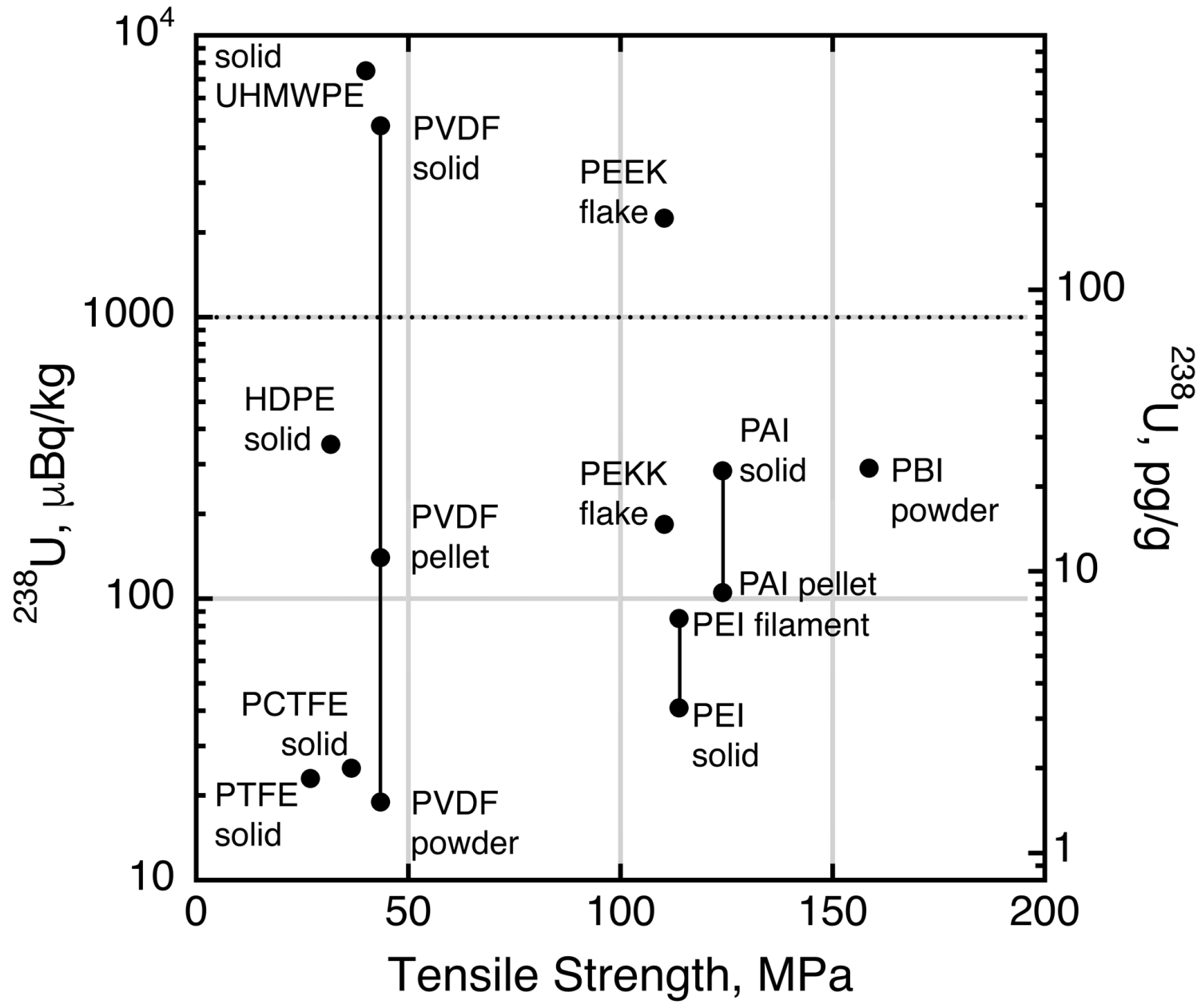

**Figure 5**. Average $^{238}U$ assay values on a log scale, as a measure of radiopurity, compared to mechanical strength as determined by standard ASTM test data for tensile strength (test D638). Assay values are from this and prior papers of our own[23, 26, 44], except for PTFE which is for NXT-75 raw Teflon from DuPont as reported in reference [12]. Strength values are from online databases such as those from Boedecker.com, or from vendor data sheets.

## 3.6 Discussion

The mass spectrometric sample preparation method described here and in recent papers[23, 26], using dry ashing in ULB EF-Cu crucibles, provides a general method for assaying ultra low levels of metals in a diversity of plastics. In this paper, it was applicable to all the high performance polymers assayed, and provided quantitative results, as opposed to upper limits. With this study, the method has now been applied to fluorinated and nonfluorinated polymers, acid-soluble and -insoluble polymers, as well as standard, engineering, and high performance plastics. Through careful sourcing of commercial polymer samples, and assay by ICP-MS with internal standards, multiple high performance polymers were found with levels for $^{232}$Th and $^{238}$U that are below—and sometimes far below—one mBq/kg. These included forms of PEKK, PBI, PAI and PEI. The PEI (Ultem) obtained as both solid and 3D printing filament material had $^{238}$U below 100 μBq/kg.

In addition, because these assays can be carried out on sub-gram quantities in a matter of days, the approach can assess materials throughout the material lifecycle from raw materials to solids and parts. Should lower detection limits be required for materials of still higher radiopurity, the method is scalable; larger Cu crucibles may be prepared by cutting and folding sheets of ULB EF-Cu, and larger polymer sample sizes may be used.

**Acknowledgements**

The authors thank Mike Oliveto from Drake Plastics Ltd. Co. for supplying PAI TORLON, helping us get the T4200 grade which is free of titanium dioxide pigment, and care in getting the rod made from the same batch of Torlon as the pellets. We thank Dan Ireland of Solvay Specialty Polymers for the sample of PEEK Ketaspire. We thank Rebecca Erikson and Josef Christ at PNNL for supplying the PEI filament material. Work on ultralow background polymers was initiated with funding from PNNL Laboratory Directed Research and Development(LDRD) funds under the Ultrasensitive Nuclear Measurements Initiative (USNMI). This paper collects results on high performance polymers from this initial effort as well as subsequent funding, including the PNNL LDRD Initiative entitled Nuclear Physics, Particle Physics, Astrophysics, and Cosmology(NPAC) and work funded by the U.S. Department of Energy (DOE) Office of Science – Nuclear Physics. The Pacific Northwest National Laboratory is a multi-program national laboratory operated for the DOE by Battelle Memorial Institute under contract number DE-AC05-76RL01830.

**References**

. .

[1] J.C. Loach, J. Cooley, G.A. Cox, Z. Li, K.D. Nguyen, A.W.P. Poon, **A database for storing the results of material radiopurity measurements**, *Nucl. Instrum. Methods Phys. Res. A*, 839 (2016) 6-11. 10.1016/j.nima.2016.09.036

[2] N. Abgrall, I.J. Arnquist, F.T. Avignone, H.O. Back, A.S. Barabash, F.E. Bertrand, M. Boswell, A.W. Bradley, V. Brudanin, M. Busch, et al., **The MAJORANA DEMONSTRATOR radioassay program**, *Nucl. Instrum. Methods Phys. Res. A*, 828 (2016) 22-36. 10.1016/j.nima.2016.04.070

[3] M. Agostini, M. Allardt, E. Andreotti, A.M. Bakalyarov, M. Balata, I. Barabanov, M.B. Heider, N. Barros, L. Baudis, C. Bauer, et al., **The background in the 0 nu beta beta experiment GERDA**, *Eur. Phys. J. C*, 74 (2014). ARTN 2764 10.1140/epjc/s10052-014-2764-z

[4] E. Aprile, K. Arisaka, F. Arneodo, A. Askin, L. Baudis, A. Behrens, K. Bokeloh, E. Brown, J.M.R. Cardoso, B. Choi, et al., **Material screening and selection for XENON100**, *Astropart. Phys.*, 35 (2011) 43-49. 10.1016/j.astropartphys.2011.06.001

[5] V. Álvarez, I. Bandac, A. Bettini, F.I.G.M. Borges, S. Cárcel, J. Castel, S. Cebrián, A. Cervera, C.A.N. Conde, T. Dafni, et al., **Radiopurity control in the NEXT-100 double beta decay experiment: procedures and initial measurements**, *J. Instrum.*, 8 (2013) T01002-T01002. 10.1088/1748-0221/8/01/t01002

[6] C. Arpesella, H.O. Back, M. Balata, T. Beau, G. Bellini, J. Benziger, S. Bonetti, A. Brigatti, C. Buck, B. Caccianiga, et al., **Measurements of extremely low radioactivity levels in BOREXINO**, *Astropart. Phys.*, 18 (2002) 1-25. Pii S0927-6505(01)00179-7 Doi 10.1016/S0927-6505(01)00179-7

[7] F. Aznar, J. Castel, S. Cebrian, T. Dafni, A. Diago, J.A. Garcia, J.G. Garza, H. Gomez, D. Gonzalez-Diaz, D.C. Herrera, et al., **Assessment of material radiopurity for Rare Event experiments using Micromegas**, *J. Instrum.*, 8 (2013) C11012. Artn C11012 10.1088/1748-0221/8/11/C11012

[8] D. Budjas, A.M. Gangapshev, J. Gasparro, W. Hampel, M. Heisel, G. Heusser, M. Hult, A.A. Klimenko, V.V. Kuzminov, M. Laubenstein, et al., **Gamma-ray spectrometry of ultra low levels of radioactivity within the material screening program for the GERDA experiment**, *Appl Radiat Isot*, 67 (2009) 755-758. 10.1016/j.apradiso.2009.01.019

[9] S. Cebrián, J. Pérez, I. Bandac, L. Labarga, V. Álvarez, A.I. Barrado, A. Bettini, F.I.G.M. Borges, M. Camargo, S. Cárcel, et al., **Radiopurity assessment of the tracking readout for the NEXT double beta decay experiment**, *J. Instrum.*, 10 (2015) P05006.

[10] P. Jagam, J.J. Simpson, **Measurements of Th, U and K concentrations in a variety of materials**, *Nucl. Instrum. Methods Phys. Res. A*, 324 (1993) 389-398. 10.1016/0168-9002(93)91000-d

[11] B.D. LaFerriere, T.C. Maiti, I.J. Arnquist, E.W. Hoppe, **A novel assay method for the trace determination of Th and U in copper and lead using inductively coupled plasma mass spectrometry**, *Nucl. Instrum. Methods Phys. Res. A*, 775 (2015) 93-98. 10.1016/j.nima.2014.11.052

[12] D.S. Leonard, P. Grinberg, P. Weber, E. Baussan, Z. Djurcic, G. Keefer, A. Piepke, A. Pocar, J.L. Vuilleumier, J.M. Vuilleumier, et al., **Systematic study of trace radioactive**

**impurities in candidate construction materials for EXO-200**, *Nucl. Instrum. Methods Phys. Res. A*, 591 (2008) 490-509. 10.1016/j.nima.2008.03.001
[13] S. Nisi, A. Di Vacri, M.L. Di Vacri, A. Stramenga, M. Laubenstein, **Comparison of inductively coupled mass spectrometry and ultra low-level gamma-ray spectroscopy for ultra low background material selection**, *Appl Radiat Isot*, 67 (2009) 828-832. 10.1016/j.apradiso.2009.01.021
[14] D.S. Leonard, D.J. Auty, T. Didberidze, R. Gornea, P. Grinberg, R. MacLellan, B. Methven, A. Piepke, J.L. Vuilleumier, J.B. Albert, et al., **Trace radioactive impurities in final construction materials for EXO-200**, *Nucl. Instrum. Methods Phys. Res. A*, 871 (2017) 169-179. 10.1016/j.nima.2017.04.049
[15] J. Dobson, C. Ghag, L. Manenti, **Ultra-low background mass spectrometry for rare-event searches**, *Nucl. Instrum. Methods Phys. Res. A*, 879 (2018) 25-30. 10.1016/j.nima.2017.10.014
[16] E. Aprile, J. Aalbers, F. Agostini, M. Alfonsi, F.D. Amaro, M. Anthony, F. Arneodo, P. Barrow, L. Baudis, B. Bauermeister, et al., **Material radioassay and selection for the XENON1T dark matter experiment**, *Eur. Phys. J. C*, 77 (2017) 890. ARTN 890 10.1140/epjc/s10052-017-5329-0
[17] S. Cebrián, J. Pérez, I. Bandac, L. Labarga, V. Álvarez, C.D.R. Azevedo, J.M. Benlloch-Rodríguez, F.I.G.M. Borges, A. Botas, S. Cárcel, et al., **Radiopurity assessment of the energy readout for the NEXT double beta decay experiment**, *J. Instrum.*, 12 (2017) T08003-T08003. 10.1088/1748-0221/12/08/t08003
[18] V. Álvarez, I. Bandac, A. Bettini, F.I.G.M. Borges, S. Cárcel, J. Castel, S. Cebrián, A. Cervera, C.A.N. Conde, T. Dafni, et al., **Radiopurity control in the NEXT-100 double beta decay experiment**, *AIP Conference Proceedings*, 1549 (2013) 46-49. 10.1063/1.4818073
[19] J. Boger, R.L. Hahn, J.K. Rowley, A.L. Carter, B. Hollebone, D. Kessler, I. Blevis, F. Dalnoki-Veress, A. DeKok, J. Farine, et al., **The Sudbury Neutrino Observatory**, *Nucl. Instrum. Methods Phys. Res. A*, 449 (2000) 172-207. https://doi.org/10.1016/S0168-9002(99)01469-2
[20] Radiopurity.org retrieved September 7 , 2019. Assay values for Victrex PEEK were reported as 20.53 and 11.85mBq/kg (emphasis - milli) for U-238 and Th-232 respectively.
[21] E. Lopez Asamar, **Low-threshold WIMP search at SuperCDMS**, *Nucl. Part. Phys. Proc.*, 273-275 (2016) 395-398. 10.1016/j.nuclphysbps.2015.09.057
[22] R. Agnese, A.J. Anderson, T. Aramaki, I. Arnquist, W. Baker, D. Barker, R.B. Thakur, D.A. Bauer, A. Borgland, M.A. Bowles, et al., **Projected sensitivity of the SuperCDMS SNOLAB experiment**, *Phys. Rev. D*, 95 (2017) 082002. ARTN 082002 10.1103/PhysRevD.95.082002
[23] I.J. Arnquist, E.J. Hoppe, M. Bliss, J.W. Grate, **Mass Spectrometric Determination of Uranium and Thorium in High Radiopurity Polymers Using Ultra Low Background Electroformed Copper Crucibles for Dry Ashing**, *Anal Chem*, 89 (2017) 3101-3107. 10.1021/acs.analchem.6b04854
[24] H. Abbasi, M. Antunes, J.I. Velasco, **Influence of polyamide-imide concentration on the cellular structure and thermo-mechanical properties of polyetherimide/polyamide-imide blend foams**, *Eur. Polym. J.*, 69 (2015) 273-283. 10.1016/j.eurpolymj.2015.06.014
[25] M.M. Teoh, T.S. Chung, K.Y. Wang, M.D. Guiver, **Exploring Torlon/P84 co-polyamide-imide blended hollow fibers and their chemical cross-linking modifications for**

**pervaporation dehydration of isopropanol**, *Sep. Purif. Technol.*, 61 (2008) 404-413. 10.1016/j.seppur.2007.12.002
[26] I.J. Arnquist, E.W. Hoppe, M. Bliss, K. Harouaka, M.L. di Vacri, J.W. Grate, **Mass spectrometric assay of high radiopurity solid polymer materials for parts in radiation and rare event physics detectors**, *Nucl. Instrum. Methods Phys. Res. A*, 943 (2019) 162443. ARTN 162443
10.1016/j.nima.2019.162443
[27] I.J. Arnquist, M.L.P. Thomas, J.W. Grate, M. Bliss, E.W. Hoppe, **A dry ashing assay method for the trace determination of Th and U in polymers using inductively coupled plasma mass spectrometry**, *J. Radioanal. Nucl. Chem.*, 307 (2016) 1883-1890. 10.1007/s10967-015-4343-7
[28] G.A. Uriano, C.C. Gravatt, G.H. Morrison, **The Role of Reference Materials and Reference Methods in Chemical Analysis**, *Crit. Rev. Anal. Chem.*, 6 (1977) 361-412. 10.1080/10408347708542696
[29] P. De Bièvre, **Isotope dilution mass spectrometry: what can it contribute to accuracy in trace analysis?**, *Fresenius J. Anal. Chem.*, 337 (1990) 766-771. 10.1007/bf00322250
[30] M.J.T. Milton, T.J. Quinn, **Primary methods for the measurement of amount of substance**, *Metrologia*, 38 (2001) 289-296. Doi 10.1088/0026-1394/38/4/1
[31] P. De Bièvre, **An isotope dilution mass spectrometric measurement procedure has the potential of being a very good reference measurement procedure, but is not a "definitive" one**, *Accredit. Qual. Assur.*, 15 (2010) 321-322. 10.1007/s00769-010-0657-x
[32] J. Vogl, W. Pritzkow, **Isotope Dilution Mass Spectrometry - A Primary Method of Measurement and Its Role for RM Certification**, *Mapan-Journal of Metrology Society of India*, 25 (2010) 135-164. DOI 10.1007/s12647-010-0017-7
[33] R.S. Dybczyński, **50 Years of adventures with neutron activation analysis with the special emphasis on radiochemical separations**, *J. Radioanal. Nucl. Chem.*, 303 (2014) 1067-1090. 10.1007/s10967-014-3822-6
[34] R.S. Dybczynski, B. Danko, H. Polkowska-Motrenko, Z. Samczynski, **RNAA in metrology: A highly accurate (definitive) method**, *Talanta*, 71 (2007) 529-536. 10.1016/j.talanta.2006.04.021
[35] B.G. Dawkins, F. Qin, M. Gruender, G.S. Copeland, Polybenzimidazole (PBI) high temperature polymers and blends, in: M.T. DeMeuse (Ed.) High Temperature Polymer Blends, Woodhead Publishing, 2014, pp. 174-212.
[36] Celazole® PBI U-60 SD. https://pbipolymer.com/wp-content/uploads/2019/04/CELAZOLE-U-60SD-Typical-Properties.pdf downloaded September 11, 2019.
[37] Drake PAI 4200 Product Datasheet, https://drakeplastics.com/Torlon-4200-non-reinforced-pai/ , downloaded September 11, 2019. See also https://drakeplastics.com/Torlon-4200-overview/.
[38] https://www.solvay.com/en/brands/torlon-pai, samples Septermber 11, 2019. See also Torlon 4200 Technical Data Sheet (Solvay), https://catalog.ides.com/Datasheet.aspx?I=92041&FMT=PDF&U=0&CULTURE=en-US&E=137213 , downloaded Septermber 11, 2019.
[39] https://www.3dxtech.com/thermax-pei-3d-filament-made-using-ultem-1010/, sampled November 19, 2019.

[40] C.E. Aalseth, A.R. Day, E.W. Hoppe, T.W. Hossbach, B.J. Hyronimus, M.E. Keillor, K.E. Litke, E.E. Mintzer, A. Seifert, G.A. Warren, **Design and construction of a low-background, internal-source proportional counter**, *J. Radioanal. Nucl. Chem.*, 282 (2009) 233-237. 10.1007/s10967-009-0258-5
[41] C.E. Aalseth, A.R. Day, E.S. Fuller, E.W. Hoppe, M.E. Keillor, B. Leferriere, E.K. Mace, J. Merriman, A.W. Myers, C.T. Overman, et al., **A new shallow underground gas-proportional counting lab--first results and Ar-37 sensitivity**, *Appl Radiat Isot*, 81 (2013) 151-155. 10.1016/j.apradiso.2013.03.050
[42] G. Zuzel, H. Simgen, **High sensitivity radon emanation measurements**, *Appl Radiat Isot*, 67 (2009) 889-893. 10.1016/j.apradiso.2009.01.052
[43] J.E. Fast, C.E. Aalseth, D.M. Asner, C.A. Bonebrake, A.R. Day, K.E. Dorow, E.S. Fuller, B.D. Glasgow, T.W. Hossbach, B.J. Hyronimus, et al., **The Multi-sensor Airborne Radiation Survey (MARS) instrument**, *Nucl. Instrum. Methods Phys. Res. A*, 698 (2013) 152-167. 10.1016/j.nima.2012.09.029
[44] E.W. Hoppe, N.R. Overman, B.D. LaFerriere, **Evaluation of ultra-low background materials for uranium and thorium using ICP-MS**, *AIP Conference Proceedings*, 1549 (2013) 58-65. 10.1063/1.4818076